# AN IMPROVED ESTIMATOR FOR POPULATION MEAN USING AUXILIARY INFORMATION IN STRATIFIED RANDOM SAMPLING

SACHIN MALIK, VIPLAV K. SINGH AND † RAJESH SINGH

*Department of Statistics, Banaras Hindu University*

*Varanasi-221005, India*

*†Corresponding author*

(sachinkurava999@gmail.com, viplavkumarsingh0802@gmail.com, rsinghstat@gmail.com)

## ABSTRACT

In the present study, we propose a new estimator for population mean $\bar{Y}$ of the study variable $y$ in the case of stratified random sampling using the information based on auxiliary variable $x$. Expression for the mean squared error (MSE) of the proposed estimators is derived up to the first order of approximation. The theoretical conditions have also been verified by a numerical example. An empirical study is carried out to show the efficiency of the suggested estimator over sample mean estimator, usual separate ratio, separate product estimator and other proposed estimator's.

***Key words***: Study variable, auxiliary variable, stratified random sampling, separate ratio estimator, bias and mean squared error.

## 1. INTRODUCTION

The problem of estimating the population mean in the presence of an auxiliary variable has been widely discussed in finite population sampling literature. Out of many ratio, product and regression methods of estimation are good examples in this context. Diana (1993) suggested a class of estimators of the population mean using one auxiliary variable in the stratified random sampling and examined the MSE of the estimators up to the $k^{th}$ order of approximation. Kadilar and Cingi (2003), Singh et al. (2007), Singh and Vishwakarma (2008), Koyuncu and Kadilar (2009) proposed estimators in stratified random sampling. Bahl and Tuteja (1991 ) and Singh et al. (2007) suggested some exponential ratio type estimators. Consider a finite population of size N and is divided into L strata such that $\sum_{h=1}^{L} N_h = N$ where $N_h$ is the size of $h^{th}$ stratum (h=1,2,...,L). We select a sample of size $n_h$ from each stratum by simple random sample without replacement (SRSWOR) sampling such that $\sum_{h=1}^{L} n_h = n$, where $n_h$ is the stratum sample size. Let ($y_{hi}$, $x_{hi}$, $z_{hi}$) denote the observed values of y, x, and z on the $i_{th}$ unit of the $h_{th}$ stratum, where i=1, 2, 3…$N_h$.

These are some notations used:

$$\bar{y}_{st} = \sum_{h=1}^{L} w_h \bar{y}_h \ , \ \bar{y}_h = \frac{1}{n_h}\sum_{i=1}^{n_h} \bar{y}_{hi} \ , \ \bar{Y}_h = \frac{1}{N_h}\sum_{i=1}^{n_h} \bar{Y}_{hi} \, ,$$

$$Y = \bar{Y}_{st} = \sum_{h=1}^{L} w_h \bar{Y}_h \, , \quad w_h = \frac{N_h}{N}.$$

Let

$$S_{yh}^2 = \sum_{i=1}^{N_h} \frac{\left(\overline{y_h} - \bar{Y}_h\right)^2}{N_h - 1}, \qquad S_{xh}^2 = \sum_{i=1}^{N_h} \frac{\left(\bar{x}_h - \bar{X}_h\right)^2}{N_h - 1}$$

$$S_{yxh} = \sum_{i=1}^{N_h} \frac{\left(\bar{x}_h - \bar{X}_h\right)\left(\bar{y}_h - \bar{Y}_h\right)}{N_h - 1} \ \text{And,} \ f_h = \frac{1}{n_h} - \frac{1}{N_h}$$

## 2. ESTIMATORS IN LITERATURE

When the population mean $\bar{X}_h$ of the hth stratum of the auxiliary variable x is known then the usual separate ratio and product estimators for population mean Y are respectively given as

$$t_1 = \sum_{h=1}^{L} w_h \bar{y}_h \left( \frac{\bar{X}_h}{\bar{x}_h} \right) \tag{2.1}$$

$$t_2 = \sum_{h=1}^{L} w_h \bar{y}_h \left( \frac{\bar{x}_h}{\bar{X}_h} \right) \tag{2.2}$$

Following Bahl and Tuteja (1991), we propose the following ratio and product exponential estimators

$$t_3 = \sum_{h=1}^{L} w_h \bar{y}_h \exp\left( \frac{\bar{X}_h - \bar{x}_h}{\bar{X}_h + \bar{x}_h} \right) \tag{2.3}$$

$$t_4 = \sum_{h=1}^{L} w_h \bar{y}_h \exp\left( \frac{\bar{x}_h - \bar{X}_h}{\bar{x}_h + \bar{X}_h} \right) \tag{2.4}$$

The MSE's of these estimators are respectively, given by

$$MSE(t_1) = \sum_{h=1}^{L} W_h^2 f_h \left[ S_{yh}^2 + R_h^2 S_{xh}^2 - 2R_h S_{yxh} \right] \tag{2.5}$$

$$MSE(t_2) = \sum_{h=1}^{L} W_h^2 f_h \left[ S_{yh}^2 + R_h^2 S_{xh}^2 + 2R_h S_{yxh} \right] \tag{2.6}$$

$$MSE(t_3) = \sum_{h=1}^{L} W_h^2 f_h \left[ S_{yh}^2 + \frac{R_h^2}{4} S_{xh}^2 - R_h S_{yxh} \right] \tag{2.7}$$

$$MSE(t_4) = \sum_{h=1}^{L} W_h^2 f_h \left[ S_{yh}^2 + \frac{R_h^2}{4} S_{xh}^2 + R_h S_{yxh} \right] \tag{2.8}$$

The usual regression estimator of population mean $\bar{Y}$ is given by

$$t_{lr} = \sum_{h=1}^{L} w_h \left[ \bar{y}_h + b_h \left( \bar{X}_h - \bar{x}_h \right) \right] \tag{2.9}$$

The MSE of the regression estimator is given by

$$var(t_{lr}) = \sum_{h=1}^{L} W_h^2 f_h S_{yh}^2 \left( 1 - \rho_h^2 \right) \tag{2.10}$$

The variance of the usual sample mean estimator $\bar{y}_h$ is given as

$$\mathrm{var}(\bar{y}_{st}) = \sum_{h=1}^{L} W_h^2 f_h S_{yh}^2 \tag{2.11}$$

Yadav et al. (2011) proposed a exponential ratio-type estimator for estimating $\bar{Y}$ as

$$t_R = \sum_{h=1}^{L} w_h \bar{y}_h \exp\left( \frac{\bar{X}_h - \bar{x}_h}{\bar{X}_h + (a_h - 1)\bar{x}_h} \right) \tag{2.12}$$

The MSE of the estimator $t_R$ is given by

$$MSE(t_R) = \sum_{h=1}^{L} W_h^2 f_h \left[ S_{yh}^2 + \frac{R_h^2}{a_h^2} S_{xh}^2 - 2\frac{R_h}{a_h} S_{yxh} \right] \tag{2.13}$$

At the optimum value of $a_h$ the MSE of the estimator $t_R$ is equal to the MSE of regression estimator $t_{lr}$ given in equation (2.9).

## 3. THE PROPOSED ESTIMATOR

Motivated by Singh and Solanki (2012), we propose an estimator of population mean $\bar{Y}$ of the study variable y as

$$t_P = \sum_{h=1}^{L} w_h \left[ \left[\lambda_1 \bar{y}_h + \lambda_2 (\bar{X}_h - \bar{x}_h)\right] \left\{ 2 - \left( \frac{\bar{X}_h}{\bar{x}_h} \right) \exp\left( \frac{\bar{X}_h - \bar{x}_h}{\bar{X}_h + \bar{x}_h} \right) \right\} \right] \tag{3.1}$$

To obtain the bias and MSE of $t_P$, we write

$$\bar{y}_{st} = \sum_{h=1}^{L} w_h \bar{y}_h = \bar{Y}(1 + e_0), \quad \bar{x}_{st} = \sum_{h=1}^{L} w_h \bar{x}_h = \bar{X}(1 + e_1)$$

Such that,

$$E(e_{0h}) = E(e_{1h}) = 0,$$

and $E(e_0^2)=\dfrac{\sum_{h=1}^{L} W_h^2 f_h S_{yh}^2}{\overline{Y}^2}$, $E(e_1^2)=\dfrac{\sum_{h=1}^{L} W_h^2 f_h S_{xh}^2}{\overline{X}^2}$, $E(e_0 e_1)=\dfrac{\sum_{h=1}^{L} W_h^2 f_h S_{yxh}^2}{\overline{Y}\,\overline{X}}$.

Expressing equation (3.1) in terms of e's, we have

$$t_P=\sum_{h=1}^{L} w_h\left\{\left[\lambda_1 \overline{Y}_h(1+e_0)-\lambda_2 \overline{X}_h\left[2-(1+e_1)^{-1}\exp\left(-\frac{e_1}{2}+\frac{e_1^2}{4}\right)\right]\right]\right\}$$

$$=\sum_{h=1}^{L} w_h\left\{\left[\lambda_1 \overline{Y}_h(1+e_0)-\lambda_2 \overline{X}_h\left[1+\frac{3e_1}{2}-\frac{15}{8}e_1^2\right]\right]\right\} \tag{3.2}$$

Neglecting the terms of e's power greater than two in expression (3.2), we have

$$(t_P-\overline{Y})=\sum_{h=1}^{L} w_h\left\{\left[\left[\lambda_1 \overline{Y}_h(1+e_0)-\lambda_2 \overline{X}_h e_1+\frac{3}{2}\lambda_1 \overline{Y}_h e_1+\frac{3}{2}\lambda_1 \overline{Y}_h e_0 e_1\right.\right.\right.$$

$$\left.\left.\left.-\frac{3}{2}\lambda_2 \overline{X}_h e_1^2-\frac{15}{8}\lambda_1 \overline{Y}_h e_1^2-\overline{Y}_h\right]\right]\right\} \tag{3.3}$$

Taking expectation on both sides of (3.3), we have the bias of the estimator $t_p$ up to the first order of approximation, as

$$B(t_P)=\sum_{h=1}^{L} w_h\left\{\overline{Y}_h(\lambda_1-1)+\frac{3}{2}\lambda_1 f_h\frac{S_{yxh}}{\overline{X}_h}-\frac{3}{2}\lambda_2 f_h\frac{S_{xh}^2}{\overline{X}_h}-\frac{15}{8}\lambda_1 \overline{R}_h\frac{S_{xh}^2}{\overline{X}_h}\right\} \tag{3.4}$$

Squaring both sides of (3.3) and neglecting the terms having power greater than two, we have

$$(t_P-\overline{Y})^2=\sum_{h=1}^{L} {w_h}^2\left[\lambda_1 \overline{Y}_h(1+e_0)-\lambda_2 \overline{X}_h e_1+\frac{3}{2}\lambda_1 \overline{Y}_h e_1-\overline{Y}_h\right]^2$$

$$(t_P-\overline{Y})^2=\sum_{h=1}^{L} {w_h}^2\left[\lambda_1^2\left(\overline{Y}_h^2 e_0+\overline{Y}_h^2+\frac{9}{4}\overline{Y}_h^2 e_1^2+3\overline{Y}_h^2 e_0 e_1\right)+\lambda_2^2 \overline{X}_h^2\right.$$

$$+\bar{Y}_h^2 - 2\lambda_1 \bar{Y}_h^2 - 2\lambda_1\lambda_2 \bar{Y}_h \bar{X}_h e_0 e - 3\lambda_1\lambda_2 \bar{Y}_h \bar{X}_h e_1^2 \Big] \quad (3.5)$$

Taking expectation of both sides of (3.5), we have the mean squared error of the estimator $t_P$ up to the first order of approximation, as

$$MSE(t_P) = \lambda_1^2 P_1 + \lambda_2^2 P_2 - 2\lambda_1\lambda_2 P_3 - 3\lambda_1\lambda_2 P_4 - 2\lambda_1 \sum_{h=1}^{L} w_h^2 \bar{Y}_h^2 + \sum_{h=1}^{L} w_h^2 \bar{Y}_h^2 \quad (3.6)$$

Where,

$$\left.\begin{aligned} P_1 &= \sum_{h=1}^{L} W_h^2 f_h S_{yh}^2 + \frac{9}{4}\sum_{h=1}^{L} W_h^2 f_h R_h^2 S_{xh}^2 + \sum_{h=1}^{L} W_h^2 \bar{Y}_h^2 + 3\sum_{h=1}^{L} W_h^2 f_h R_h S_{yxh} \\ P_2 &= \sum_{h=1}^{L} W_h^2 f_h S_{xh}^2 \\ P_3 &= \sum_{h=1}^{L} W_h^2 f_h S_{yxh} \\ P_4 &= \sum_{h=1}^{L} W_h^2 f_h R_h S_{xh}^2 \end{aligned}\right\} \quad (3.7)$$

Partially differentiating expression (3.6) with respect to $\lambda_1$ and $\lambda_2$, we get the optimum values of $\lambda_1$ and $\lambda_2$ as

$$\lambda_1(opt) = \frac{4P_2 \sum_{h=1}^{L} w_h^2 \bar{Y}_h^2}{4P_1P_2 - [2P_3 + 3P_4]^2} \quad \text{and} \quad \lambda_2(opt) = \frac{2[2P_3 + 3P_4]\sum_{h=1}^{L} w_h^2 \bar{Y}_h^2}{4P_1P_2 - [2P_3 + 3P_4]^2}$$

Putting these optimum values of $\lambda_1$ and $\lambda_2$ in expression (3.7), we get the minimum value of the MSE($t_P$).

## 4. NUMERICAL STUDY

For numerical study, we use the data set earlier used by Kadilar and Cingi (2003). In this data set, Y is the apple production amount and X is the number of apple trees in 854 villages of Turkey in 1999. The population information about this data set is given in Table 4.1.

**TABLE 4.1: DATA STATISTICS**

| | | |
|---|---|---|
| N=854 | n=140 | |
| $N_1$=106 | $N_2$=106 | $N_3$=94 |
| $N_4$=171 | $N_5$=204 | $N_6$=173 |
| $n_1$=9 | $n_2$=17 | $n_3$=38 |
| $n_4$=67 | $n_5$=7 | $n_6$=2 |
| $\overline{X}_1 = 24375$ | $\overline{X}_2 = 27421$ | $\overline{X}_3 = 72409$ |
| $\overline{X}_4 = 74365$ | $\overline{X}_5 = 26441$ | $\overline{X}_6 = 9844$ |
| $\overline{Y}_1 = i536$ | $\overline{Y}_2 = 2212$ | $\overline{Y}_3 = 9384$ |
| $\overline{Y}_4 = 5588$ | $\overline{Y}_5 = 967$ | $\overline{Y}_6 = 404$ |
| $\beta_{x1} = 25.71$ | $\beta_{x2} = 34.57$ | $\beta_{x3} = 26.14$ |
| $\beta_{x4} = 97.60$ | $\beta_{x5} = 27.47$ | $\beta_{x6} = 28.10$ |
| $C_{x1}$=2.02 | $C_{x2}$=2.10 | $C_{x3}$=2.22 |
| $C_{x4}$=3.84 | $C_{x5}$=1.72 | $C_{x6}$=1.91 |
| $C_{y1}$=4.18 | $C_{y2}$=5.22 | $C_{y3}$=3.19 |
| $C_{y4}$=5.13 | $C_{y5}$=2.47 | $C_{y6}$=2.34 |
| $S_{x1}$=49189 | $S_{x2}$=57461 | $S_{x3}$=160757 |
| $S_{x4}$=285603 | $S_{x5}$=45403 | $S_{x6}$=18794 |
| $S_{y1}$=6425 | $S_{y2}$=11552 | $S_{y3}$=29907 |
| $S_{y4}$=28643 | $S_{y5}$=2390 | $S_{y6}$=946 |
| $\rho_1 = 0.82$ | $\rho_2 = 0.86$ | $\rho_3 = 0.90$ |
| $\rho_4 = 0.99$ | $\rho_5 = 0.71$ | $\rho_6 = 0.89$ |
| $f_1 = 0.102$ | $f_2 = 0.049$ | $f_3 = 0.016$ |

$f_4 = 0.009$ $f_5 = 0.138$ $f_6 = 0.006$

$w_1^2 = 0.015$ $w_2^2 = 0.015$ $w_3^2 = 0.012$

$w_4^2 = 0.04$ $w_5^2 = 0.057$ $w_6^2 = 0.041$

For the purpose of the efficiency comparison of the proposed estimator, we have computed the percent relative efficiencies (PREs) of the estimators with respect to the usual unbiased estimator $\bar{y}_{st}$ using the formula:

$$PRE\left(t, \bar{y}_{st}\right) = \frac{MSE\left(\bar{y}_{st}\right)}{MSE(t)} * 100$$, where $t = (t_1, t_2, t_3, t_{lr}, t_P)$

The findings are given in the Table 4.2.

**TABLE 4.2: PERCENT RELATIVE EFFICIENCES (PRE) OF ESTIMATORS**

| S. No. | ESIMATORS | PRE'S |
|---|---|---|
| 1 | $\bar{y}_{st}$ | 100 |
| 2 | $t_1$ | 423.20 |
| 3 | $t_2$ | 37.60 |
| 2 | $t_3$ | 199.14 |
| 3 | $t_4$ | 12.83 |
| 4 | $t_{lr}$ | 629.03 |
| 5 | $t_R$ | 629.03 |
| 6 | $t_P$ | 789.87 |

## 6. CONCLUSION

In this paper, we have proposed a new estimator for estimating unknown population mean of study variable using auxiliary variables. Expressions for bias and MSE of the estimator are derived up to first order of approximation. The proposed estimator is compared with usual mean estimator and other considered estimators. A numerical study is carried out to support the theoretical results. From Table 4.2, it is clear that the proposed estimator $t_P$ is more efficient than the unbiased sample mean estimator $\bar{y}_{st}$, usual ratio and product estimator $t_1$ and $t_2$, usual exponential ratio and product type estimators $t_3$ and $t_4$ and Yadav et al. (2011) estimator $t_R$.

**Acknowledgement :** The authors are thankful to the referee's for their valuable comments and suggestions regarding the improvement of the paper.

## REFERENCES

Bahl, S. and Tuteja, R.K. (1991): Ratio and product type exponential estimator. Infrm. and Optim. Sci., XII, I, 159-163.

Diana, G. (1993). A class of estimators of the population mean in stratified random sampling. *Statistica* 53(1):59–66.

Kadilar,C. and Cingi,H. (2003): Ratio Estimators in Straitified Random Sampling. Biometrical Journal 45 (2003) 2, 218-225.

Koyuncu, N. and Kadilar, C. (2009) : Family of estimators of population mean using two auxiliary variables in stratified random sampling. Comm. In Stat. – Theory and Meth., 38:14, 2398-2417.

Solanki, R.S., Singh, H.P., Rathour, A.(2012): An alternative estimator for estimating the finite population mean using auxiliary information in sample surveys. ISRN Prob. and Stat. doi:10.5402/2012/657682

Singh, H., P. and Vishwakarma, G. K. (2008): A family of estimators of population mean using auxiliary information in stratified sampling. Communication in Statistics Theory and Methods, 37(7), 1038-1050.

Singh, R., Chauhan, P., Sawan, N. and Smarandache,F. (2007): Auxiliary information and a priori values in construction of improved estimators. Renaissance High press.

Yadav R., Upadhyaya L.N., Singh H.P., Chatterjee S. (2011): Improved separate exponential estimator for population mean using auxiliary information. Statistics in Transition_new series, 12(2), pp. 401-412.